\documentclass{sig-alternate-05-2015}
\usepackage{multirow}
\usepackage{url}
\begin{document}

\setcopyright{rightsretained}

\doi{10.475/123_4}

\isbn{123-4567-24-567/08/06}



%

\title{Voting with Feet: Who are Leaving Hillary Clinton and Donald Trump?}

%
%
%
%
%

\numberofauthors{6} 
%
\author{
%
%
\alignauthor
{Yu Wang}\\
	   \affaddr{Political Science}\\
       \affaddr{University of Rochester}\\
       \affaddr{Rochester, NY, 14627}\\
       \email{ywang@ur.rochester.edu}
       \alignauthor
Yuncheng Li\\
	   \affaddr{Computer Science}\\
       \affaddr{University of Rochester}\\
       \affaddr{Rochester, NY, 14627}\\
       \email{yli@cs.rochester.edu}   \\
         \alignauthor
       Quanzeng You\\
       \affaddr{Computer Science}\\
       \affaddr{University of Rochester}\\
       \affaddr{Rochester, NY, 14627}\\
       \email{qyou@cs.rochester.edu}
       \and  
       \alignauthor        
       Xiyang Zhang\\
	   \affaddr{Psychology}\\
       \affaddr{University of Oklahoma}\\
       \affaddr{Norman, OK, 73019}\\
       \email{xiyang.zhang-1@ou.edu}
       \alignauthor   
       Richard Niemi\\
	   \affaddr{Political Science}\\
       \affaddr{University of Rochester}\\
       \affaddr{Rochester, NY, 14627}\\
       \email{niemi@rochester.edu}      
         \alignauthor
         Jiebo Luo\\
       \affaddr{Computer Science}\\
       \affaddr{University of Rochester}\\
       \affaddr{Rochester, NY, 14627}\\
       \email{jluo@cs.rochester.edu}  
}

\maketitle
\begin{abstract}

From a crowded field with 17 candidates, Hillary Clinton and Donald Trump have emerged as the two front-runners in the 2016 U.S. presidential campaign. The two candidates each boast more than 5 million followers on Twitter, and at the same time both have witnessed hundreds of thousands of people leave their camps. In this paper we attempt to characterize individuals who have left Hillary Clinton and Donald Trump between September 2015 and March 2016.

Our study focuses on three dimensions of social demographics: social capital, gender, and age. Within each camp, we compare the characteristics of the current followers with former followers, i.e., individuals who have left since September 2015. We use the number of followers to measure social capital, and profile images to infer gender and age. For classifying gender, we train a convolutional neural network (CNN). For age, we use the Face++ API.

Our study shows that for both candidates followers with more social capital are more likely to leave (or switch camps). For both candidates females make up a larger presence among unfollowers than among current followers. Somewhat surprisingly, the effect is particularly pronounced for Clinton. Lastly, middle-aged individuals are more likely to leave Trump, and the young are more likely to leave Hillary Clinton.

\end{abstract}

%
%



\section{Introduction}
\begin{figure}[!h]
\centering
\includegraphics[height=5cm,width=8.4cm]{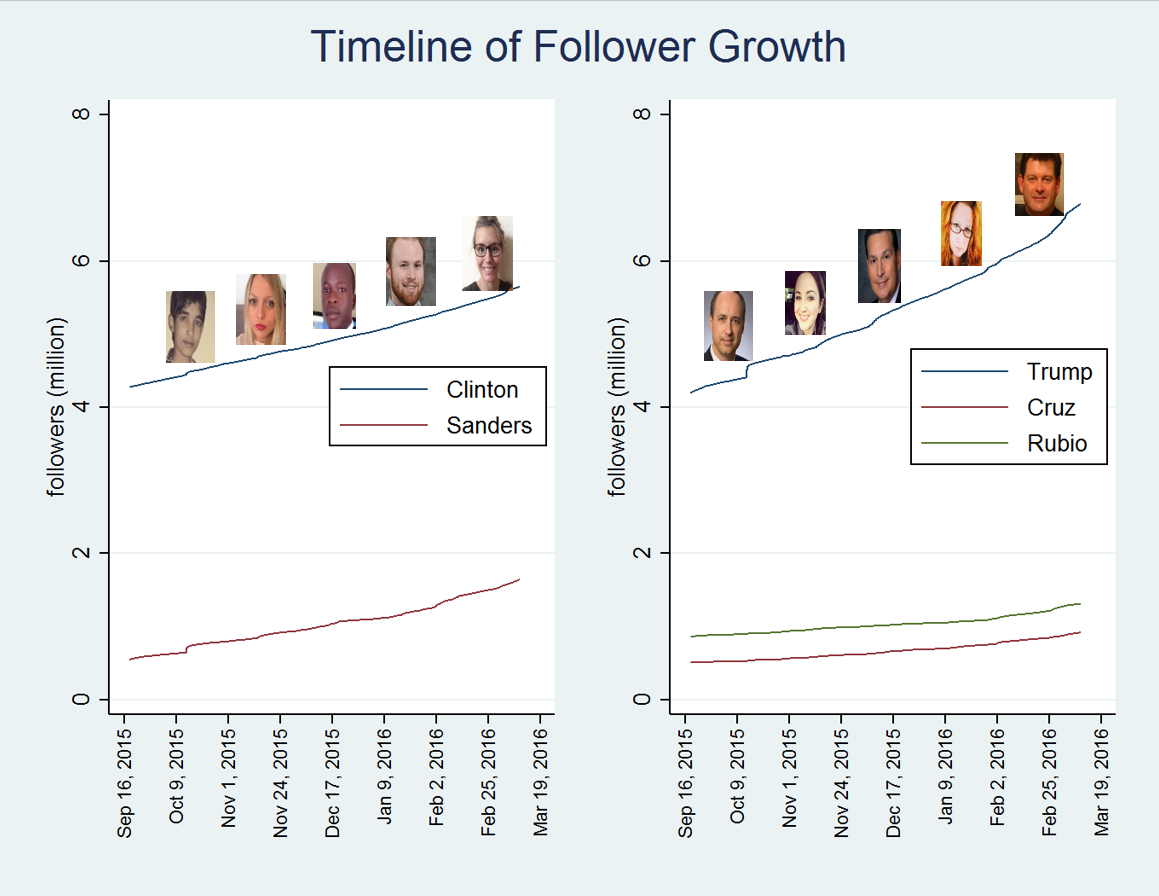}
\caption{Donald Trump and Hillary Clinton lead in terms of Twitter followers.}
\label{follower-timeline}
\end{figure}

From a crowded field with 17 candidates in September 2015, Hillary Clinton and Donald Trump have emerged as the two front-runners in the 2016 U.S. presidential campaign. The two candidates rely considerably on Twitter to reach out to voters, disseminate information and attack rival candidates. Between September 18th 2015 to March 1st 2016 (the first Super Tuesday), Hillary Clinton posted 1973 tweets and Donald Trump posted 3175. Both candidates lead in terms of Twitter followers: Clinton has 5.8 million followers and Trump has 7.3 million (Figure \ref{follower-timeline}), and both see large turnovers in their followers. In this paper we attempt to characterize these individuals who have left Hillary Clinton and Donald Trump.

Our study focuses on three dimensions of social demographics: social capital, gender, and age. Within each camp, we compare the characteristics of those who have left (i.e. unfollowers) with the current followers. We use the number of followers to measure social capital, and use profile images to infer gender and age. For classifying gender, we train a convolutional neural network (CNN). For age, we use the Face++ API.\footnote{http://www.faceplusplus.com/.}




Our study finds that those who have left Clinton are predominantly female, young, and socially well positioned. A considerable portion of them are now in the Bernie Sanders camp. Those who have left Donald Trump tend to be middle-aged and socially well positioned. Trump unfollowers are also more likely to be female, but the effect is \textit{not} as pronounced as that for Hillary Clinton. Rather than switch to Ted Cruz or John Kasich, Trump unfollowers are actually more likely to be following the Democratic candidates.



\section{Related Work}



Our work builds upon previous research in electoral studies and in computer vision.

In eletoral studies, researchers have found that gender constitutes an important factor in voting behavior. One common observation is that women tend to vote for women, which is usually referred to as gender affinity effect \cite{sexAndGOP,genderAffinityEffect}. In our work, we will test compared with her currently followers whether women are less likely to unfollow Clinton, i.e. the retention rate is high. 


In data mining, there is a burgeoning literature on using social media data to analyze and predict elections. In particular, several studies have explored ways to infer users' preferences. According to \cite{tweets2polls}, tweets with sentiment can potentially serve as votes and substitute traditional polling. \cite{trumponfire} exploits the variations in the number of `likes' of the tweets to infer Trump followers' topic preferences. \cite{neco} assumes that more followers will translate into more votes and uses follower growth on public debate dates to measure candidates' debate performance. Our work, by contrast, explores the unique preference-revealing behavior of unfollowing.




Our work ties in with current computer vision research, as the profile images of the followers constitute an integral part of our \textit{US2016} dataset. In this dimension, our work is related to gender and age classification using facial features. \cite{israel} uses a five-layer network to classify both age and gender.  \cite{trumpists} uses user profile images to study and compare the social demographics of Trump followers and Clinton followers, and is most closely related to our work. We improve on their work by analyzing both followers and unfollowers and comparing the differences of the two groups. Furthermore, our work is able to track which candidates these unfollowers have switched to.

\cite{trumponfire} employs LDA to model tweet topics and use negative binomial regression on the number of tweet `likes' to infer topic preferences of Trump followers. Our work, by comparison, explores another aspect of preference-revelation: the actions of unfollowing.

\section{Data and Methodology}

In this section, we describe our dataset \textit{US2016}, the pre-processing procedures and our CNN model. One variable is \textit{number of followers.} This variable is available for both candidates and covers the entire period from Sept. 18, 2015 to March 1, 2016 (the first Super Tuesday). From a sample of 3,289,271  Trump followers in September 2015, we identified 188,507 individuals who have left Trump by March 2016. This makes an unfollowing rate of 5.73\%.  From a sample of 3,585,000 Clinton followers in September 2015, we identified 133,694 individuals who have left Clinton by March 2016. This makes a leaving rate of 3.73\%. In Tables 1 and 2, we report the identified destinations of these unfollowers.

\begin{table}[!h]
\setlength{\tabcolsep}{3.5pt}
\renewcommand{\arraystretch}{1}
\centering
\caption{Mobility of Clinton Followers}
\label{}
\begin{tabular}{llll}
\hline\hline
\multicolumn{4}{l}{Destination of Clinton Unfollowers} \\
\hline
Bernie Sanders  & Donald Trump & Ted Cruz & Marco Rubio \\
13.32\%         & 6.68\%       & 1.60\%   & 2.04\%   \\
\hline
\end{tabular}
\end{table}

\begin{table}[!h]
\setlength{\tabcolsep}{3.5pt}
\renewcommand{\arraystretch}{1}
\centering
\caption{Mobility of Trump Followers}
\label{}
\begin{tabular}{llll}
\hline\hline
\multicolumn{4}{l}{Destination of Former Trump Followers} \\
\hline
Hillary Clinton & Bernie Sanders & Ted Cruz & Marco Rubio \\
6.72\%          & 5.09\%         & 4.50\%   & 4.53\%   \\  
\hline
\end{tabular}
\end{table}

For each candidate's followers we also have data on user name, number of followers, geographical information and profile images. Here we make the same assumption as in \cite{trumpists} and use the number of followers as a proxy for social capital, assuming that individuals with a larger number of followers possess more social capital. We derive from the profile images the followers' demographic information, such as age and gender.

To process the profile images, we first use OpenCV to identify faces, as the majority of profile images only contain a face.\footnote{http://opencv.org/.} We discard images that do not contain a face and the ones in which OpenCV is not able to detect a face. When multiple faces are available, we choose the first one. Out of all facial images thus obtained, we select only the large ones. Here we set the threshold to 25kb. This ensures high image quality and also helps remove empty faces. Lastly we resize those images to (28, 28). In Table \ref{image}, we report the summary statistics of the images in \textit{US2016}.

\begin{table}[!h]
\centering
\caption{Number of Profile Images in 	\textit{US2016}}
\setlength{\tabcolsep}{11pt}
\label{image}
\begin{tabular}{lll}
\hline\hline
            & Hillary Clinton & Donald Trump \\
            \hline
Followers   & 21,699             & 36,937          \\
Unfollowers & 25,350             & 36,907     \\
\hline    
\end{tabular}
\end{table}

\begin{figure*}[!htb]
\centering
\includegraphics[height=5cm,width=14cm]{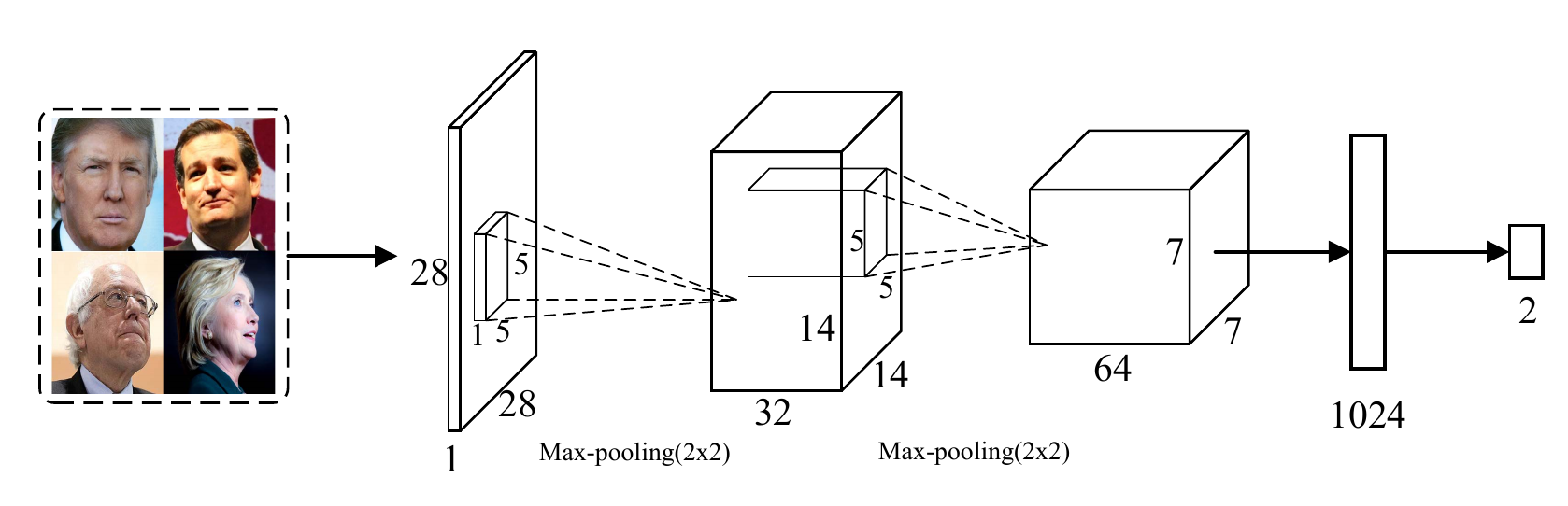}
\caption{The CNN model consists of 2 convolution layers, 2 max-pool layers and a fully connected layer.}
\label{visio}
\end{figure*}

We use a labeled data set of 2,000 Twitter images for gender classification.\footnote{The data set is now available at the first author's website.} We randomly select 1600 images for training and use the remaining 400 for validation. The architecture of our convolutional neural network is reported in Figure \ref{visio}, and the performance of the model is reported in Table \ref{Performance}.

\begin{table}[!h]
\centering
\caption{Summary Statistics of CNN Performance}
\label{Performance}
\setlength{\tabcolsep}{6.5pt}
\begin{tabular}{lllll}
\hline\hline
Architecture & Precision & Recall & F1    & Accuracy \\
2CONV-1FC    & 0.856     & 0.856  & 0.856 & 0.845   \\
\hline
\end{tabular}
\end{table}

For lack of data to train the multiple age classes, we decide that for the purpose of age classification, we use the software service from Face++.

\section{Main Results}
In this section, we present our main results. We first report on social capital, then on gender and thirdly on age. For each dimension, we present results on Clinton and Trump in parallel.

\subsection{Social Capital}
\begin{figure}[!h]
\includegraphics[height=3cm,width=8.4cm]{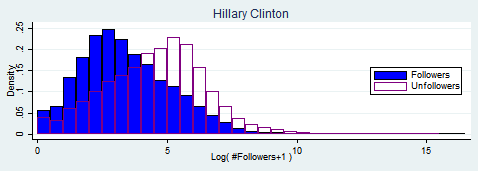}
\caption{Individuals who have left Clinton tend to have more followers themselves.}
\label{clinton-status}
\end{figure}

\begin{figure}[!h]
\includegraphics[height=3cm,width=8.4cm]{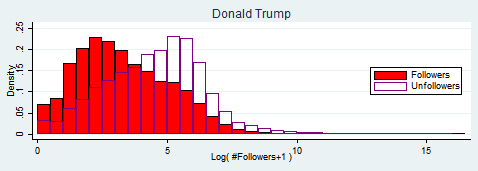}
\caption{Individuals who have left Trump tend to have more followers themselves.}
\label{trump-status}
\end{figure}

In this subsection, we analyze the distributions of the candidates' followers in terms of their own number of followers. For both Hillary Clinton and Donald Trump, we compare the distributions of the unfollowers we identified between September 2015 and March 2016 with the current followers. We present the results in Figure \ref{clinton-status} and Figure \ref{trump-status}.

For both candidates, we find that it is those who themselves have more followers that are more likely to leave.

\subsection{Gender}
In this subsection, we analyze the gender composition of Clinton's and Trump's unfollowers.

A number of studies have demonstrated the ``gender affinity effect'' in American elections \cite{women4women,runAsWomen}. In the first Democratic public debate in October, 2015, Hillary Clinton emphasized her identity as a woman: ``Being the first woman president would be quite a change from the presidents we've had, including President Obama.''\footnote{http://www.huffingtonpost.com/entry/hillary-clinton-first-woman-president\_561dbf71e4b028dd7ea5af6c.} Clinton also enjoys the support of her fellow female politicians. Out of 14 female Democratic senators, 13 have endorsed Clinton's presidential campaign.\footnote{http://www.cnn.com/2015/11/30/politics/hillary-clinton-elizabeth-warren-fundraiser.} But there is also strong evidence that Clinton's support among average Democratic women has fallen sharply.\footnote{https://www.washingtonpost.com/politics/poll-sharp-erosion-in-clinton-support-among-democratic-women/2015/09/14/6406e2a0-58c3-11e5-b8c9-944725fcd3b9\_story.html.} This makes our gender analysis of the Clinton unfollowers particularly interesting.\\

\begin{figure}[!h]
\includegraphics[height=3cm,width=8.4cm]{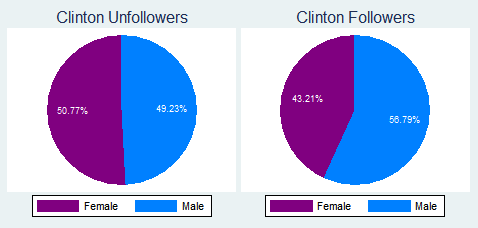}
\caption{Compared with current followers, Clinton's unfollowers are more likely to be female.}
\label{followers}
\end{figure} 

\begin{figure}[!h]
\includegraphics[height=3cm,width=8.4cm]{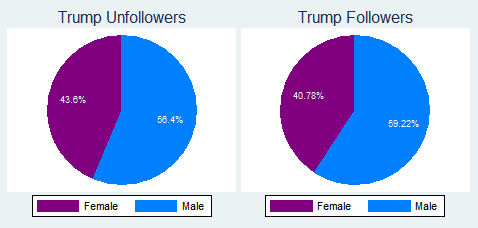}
\caption{Compared with current followers, Trump's unfollowers are more likely to be female. The effect is \textit{less} pronounced than Clinton.}
\label{followers}
\end{figure} 

The main reading is that we have not detected any gender affinity effect among the Clinton unfollowers. Females make up a larger proportion among Clinton's unfollowers than among her followers. The same holds for Donald Trump, but the effect is not nearly as pronounced as that for Clinton. This is in spite of Trump's feuds with Megyn Kelly and Carly Fiorina.

Using score test (Table \ref{stgender}), we are able to show that for both Clinton and Trump their unfollowers are statistically more likely to be female than their current followers.

\begin{table}[h!]
\centering
\caption{Score Test on Gender Composition}\label{stgender}
\setlength{\tabcolsep}{3pt}
\begin{tabular}{lllll}\hline\hline
\multirow{2}{*}{Null Hypothesis}& \multicolumn{2}{c}{Clinton} & \multicolumn{2}{c}{Trump} \\
\cline{2-3}\cline{4-5}
                                 & z statistic     & \textit{p}  value       & z statistic     & \textit{p}  value \\\hline
p$_{unfollow}$=p$_{follow}$      & 19.04           & 0                         & 7.74 & 0   \\\hline
\end{tabular}
\end{table}

\subsection{Age}
In this subsection, we study the age distribution of the candidates' unfollowers. In particular, we are interested in answering which age group is more likely to leave as compared with current followers.

\begin{figure}[!h]
\includegraphics[height=4cm,width=8.4cm]{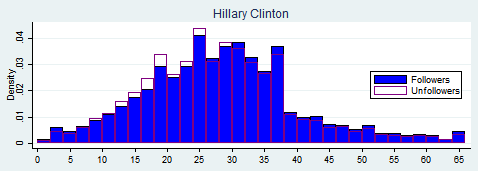}
\caption{Compared with current followers, Clinton's unfollowers are more likely to be between 12 and 26.}
\label{clinton-age}
\end{figure}

We address this demographic question by comparing the age distribution of the candidates' unfollowers with the current followers. We report our results in Figure \ref{clinton-age}. Consistent with real world voting, we find that people between 12 and 26 are more likely to leave the Clinton camp.\footnote{See, for example, \url{http://www.slate.com/articles/news_and_politics/politics/2016/02/hillary_clinton_is_losing_young_voters_to_bernie_sanders.html.}} We also find that individuals aged between 27 and 42 occupy a larger presence among the Trump unfollowers than among the Trump followers.

Using score test (Table \ref{stage}), we are able to show that the Clinton unfollowers are statistically more likely to be in the 12-26 age group than Clinton followers and that the Trump unfollowers are more likely to be 27-42 age group.

\begin{figure}[!h]
\includegraphics[height=4cm,width=8.4cm]{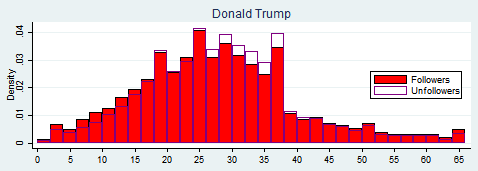}
\caption{Compared with current followers, Trump's unfollowers are more likely to be between 27 and 42.}
\label{followers}
\end{figure} 

\begin{table}[h!]
\centering
\caption{Score Test on Age Composition}\label{stage}
\setlength{\tabcolsep}{3pt}
\begin{tabular}{lllll}\hline\hline
\multirow{2}{*}{Null Hypothesis}& \multicolumn{2}{c}{Clinton (12-26)} & \multicolumn{2}{c}{Trump (27-42)} \\
\cline{2-3}\cline{4-5}
                                 & z statistic     & \textit{p}  value       & z statistic     & \textit{p}  value \\\hline
p$_{unfollow}$=p$_{follow}$      & 12.30           & 0                         & 15.69 & 0  \\\hline
\end{tabular}
\end{table}

\section{Conclusion}
We have studied the social demographics of the two leading presidential candidates' followers and unfollowers. For each candidate, we compared the characteristics of the current followers with the former followers. Our study has focused on three dimensions of social demographics: social capital, gender, and age. 

Our study shows that for both candidates followers with more social capital are more likely to leave. Also, the unfollowers are more likely to be female than the followers. The phenomenon is particularly pronounced for Clinton. Lastly, middle-aged individuals are more likely to leave Trump and the young are more likely to leave Clinton.

It is important to note that our study is based on the actual following and unfollowing actions of high-potential voters at a very large scale. This is akin to voting with their feet, thus is arguably more reliable than polling data. 


\bibliographystyle{abbrv}
\bibliography{yu}  
%
%

\end{document}